# ORBIT FEEDBACK USING X-RAY BEAM POSITION MONITORING AT THE ADVANCED PHOTON SOURCE

Glenn Decker, Om Singh, Argonne National Laboratory, Argonne, IL 60439, USA

Abstract
   The Advanced Photon Source (APS) was commissioned in 1995 as a third-generation x-ray user facility. At that time orbit control was performed exclusively with broadband rf beam position monitors (BPMs). Since then, emphasis has been placed on incorporating x-ray beam position monitors into the orbit control algorithms. This has resulted in an order of magnitude improvement in long-term beam stability vertically, using x-ray BPMs (X-BPMs) on bending magnet beamlines. Additional processing will allow similar improvements horizontally, once systematic effects associated with variable insertion device (ID) x-ray beams are properly compensated. Progress to date and upgrade plans will be presented, with an emphasis on the details of the required digital signal processing.

## 1 INTRODUCTION

The APS is an x-ray synchrotron radiation user facility based on a 7-GeV electron storage ring. Highly collimated synchrotron radiation x-rays are emitted from the electron beam as it traverses bending magnet and IDs. A primary performance measure is the allowable amount of particle beam motion (and thus x-ray beam motion).

Table 1: APS Beam Stability Specification

|  | Horizontal (rms) | Vertical (rms) |
|---|---|---|
| **Initial Design Values** | 17 microns | 4.5 microns |
| **Low Emittance / 1% Coupling Values** | 12.6 microns / 900 nanorad. | 0.59 microns / 120 nanorad. |

Shown in Table 1 are beam stability specifications for the APS. As originally stated in the APS initial design circa 1991, they are incomplete insofar as no explicit angular pointing stability specification is given. Further, both the initial values and the present low-emittance values are incomplete since no frequency band or time scale is indicated. The low-emittance mode of operation has smaller horizontal and vertical particle beam sizes in comparison to the initial design. The values in Table 1 are generally understood to apply to frequencies up to 30 Hz and down as low as reasonably achievable. Typically, APS users are most interested in beam position drifts over minutes to hours, fill-to-fill reproducibility (24 hours), and run-to-run reproducibility (months). To get an idea of scale, 100 nanoradians is the angle subtended by an object 1 mm high, observed from a distance of 10 kilometers.

## 2 SYSTEM DESCRIPTION

Excellent descriptions of the APS orbit correction systems are available in the literature [1,2,3,4]. Key elements include 360 broadband rf BPMs, 48 narrowband rf BPMs, 38 bending magnet x-ray BPMs and 48 ID x-ray BPMs. Data from these monitors is used to control up to 317 combined-function horizontal/vertical corrector magnets. Of the 317 correctors, 38 are mounted on thin-walled vacuum spool pieces allowing, in principle, correction bandwidth up to 200 Hz. The balance of the correctors is otherwise identical, but mounted around the standard thick-walled aluminum vacuum chamber. Eddy currents in the aluminum limit their bandwidth to less than 10 Hz.

X-ray BPMs are constructed of metallized, chemical-vapor-deposited (CVD) diamond wafers placed edge on to the x-ray beam [5]. Photocurrents are measured for blades placed above and below the beam for bending magnet beams and arrayed symmetrically in sets of four for collimated ID beams (Figure 1).

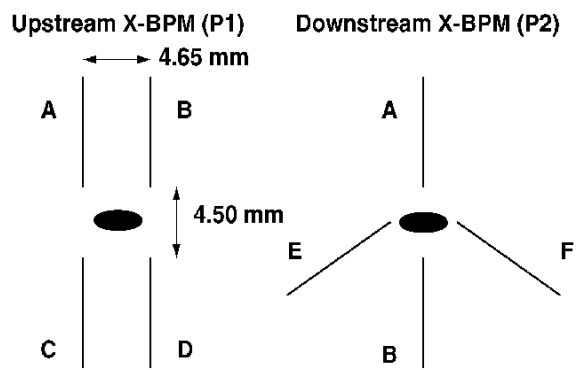

Figure 1. APS insertion device X-BPM geometry.

A pair of x-ray BPMs is mounted just inside the accelerator enclosure in each beamline "front end," connecting the accelerator vacuum to the beamline located outside of the enclosure, on the experiment hall floor. Each x-ray BPM is mounted on an X-Y translation stage with resolution better than 1 micron and better than 10-micron reproducibility. Software that uses these translation stages for convenient device calibration is in place. Two portable translation stage IOC crates have been developed for periodic calibrations and maintenance. Data acquisition for the x-ray BPM photocurrent signals has recently been completed for all installed units [1].

Data acquisition for the broadband rf BPMs takes place in customized VXI modules with simple boxcar averagers generating EPICS process variable (PV) values. A separate fiber optic communications link provides data to the real-time feedback system residing in a nearby VME input/output controller (IOC) crate (21 total). For the narrowband rf and x-ray BPMs, a dedicated 32-channel, 16-bit analog-to-digital converter (ADC) resides in the local real-time feedback IOC crate. Also included in the real-time feedback crates are two Pentek 4284 digital signal processing (DSP) modules, one Pentek 4283 DSP, a VMIC 5588 reflective memory module to provide real-time data transfer between crates, a 12-bit ADC for tracking auxiliary BPM status information, and an MVME 167 IOC module using a Motorola 68040 central processing unit (CPU). Also included is a power supply interface module for the real-time system.

Two of the DSPs in each real-time feedback crate perform the matrix multiplication operations connecting up to 160 broadband rf BPM readbacks to two local corrector power supplies, one horizontal and one vertical, at a 1.5-kHz rate. The third DSP is dedicated to processing data from the 16-bit ADC. Functions include digital filtering of position signals from the narrow-band rf BPMs, and computation and filtering of position signals from raw x-ray BPM blade photocurrent signals. Data is transferred to the vertical real-time feedback DSP and broadcast around the facility through the reflective memory network. This same data is deposited in dual-port RAM for transfer to the EPICS IOC module. Further filtering is performed in the IOC to provide pretty well de-aliased, approximately 1-Hz bandwidth signals as EPICS process variables.

A total of 21 real-time feedback IOC crates are located around the APS storage ring circumference, all connected by the reflective memory network (Figure 2). Each of 20 "slave" crates provides coverage for two APS sectors. The twenty-first crate has additional processing and provides supervisory and diagnostic support, including a forty-channel virtual oscilloscope, which can monitor any of the signals accessible on the reflective memory. This crate also provides process variables indicating total rms beam motion averaged over a maximum of 80 locations. These values are logged every 60 seconds and can be compared to the values in Table 1 to verify compliance with beam stability requirements.

A twenty-second crate has recently been added to the reflective memory network, known as the "data pool" IOC. Its purpose is ultimately to replace the "DC" orbit correction software that presently runs at the workstation level with a 2.5-second update rate. The real-time feedback system is limited in flexibility owing to the 160 by 38 feedback matrix size, and therefore explicitly avoids correcting DC orbit motion by using a 0.1-Hz high-pass filter. The workstation-based software in principle corrects everything up to this 0.1-Hz frequency, but in practice, a "dead band" at 0.1 Hz is seen. This is very unfortunate, since transient effects caused, for example, by ID gap closures, can persist for tens of seconds and have tens of microns in amplitude. The transients have been reduced somewhat by sending information from the DC algorithm to the real-time system [6].

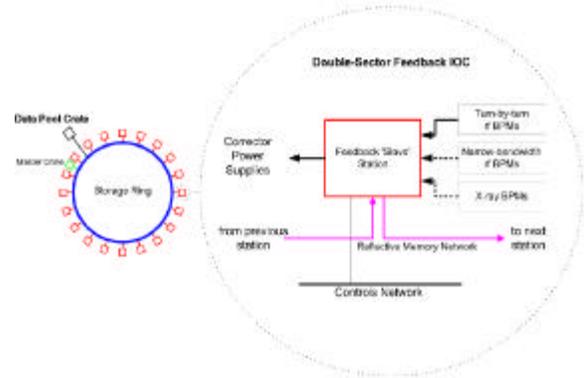

Figure 2. Layout of real-time feedback systems.

To remove the source of the transient, an ID gap feedforward solution for this has been tested using the workstation-based software. Unfortunately, because the convergence time is so slow, the generation of the feedforward lookup tables is extremely time consuming in addition to being subject to longer time scale systematic errors. By pooling all BPM data in a single IOC and running the present workstation-based software on that IOC, we hope to be able to speed up the DC orbit correction algorithm from 0.4 Hz to 50 or even 100 Hz. Feedforward algorithms need not run on this IOC, however the elimination of the 0.1-Hz dead band is essential to the determination of lookup tables for the dozens of IDs installed in the APS accelerator enclosure.

## 3 PERFORMANCE

As with any system of this complexity, performance is controlled to a large extent by the system configuration. A Tcl application is provided for this purpose, written by the APS operations analysis group [7]. This tool allows for the selection of arbitrary BPMs, correctors, and BPM weights, and allows for different accelerator configurations (e.g., low emittance vs. standard quadrupole and sextupole magnet settings). The tool uses a singular value decomposition algorithm to invert the BPM/corrector response matrix (if possible), giving the option of limiting the number of eigenvalues. Once an inverse response matrix (IRM) is available, a separate workstation-based Tcl application called controllaw uses the IRM to perform the actual orbit correction. Options like update rate, gain, despiking, etc. are available with this application at run time.

Typical "flavors" of orbit correction can roughly be categorized as being either global or local. Local correction typically involves a high density of steering correctors straddling an x-ray source point, allowing independent control of displacement and angle, without interfering with the steering of nearby beamlines. While allowing a high degree of flexibility, local steering has the disadvantage of requiring relatively large corrector changes (Amps) for each unit of steering (microns or microradians). During normal operation, local steering is generally only performed on demand from a particular beamline.

For standard operation, once local beamline steering is complete, a "global" algorithm using many more BPMs than correctors is used. This has the advantage of a large corrector "lever arm," in addition to being relatively insensitive to systematic errors affecting any individual BPM. In effect, a smooth curve is arrived at providing the best fit to the available BPM data. At the APS generally two correctors are used in each of the forty sectors. For the horizontal plane, only rf BPMs are presently in use (ID x-ray BPMs await the availability of the data pool IOC and feedforward). As many BPMs as possible are used. In the vertical plane, the broadband rf BPMs exhibit a hypersensitivity to bunch fill pattern [8] and are not used during top-up operation (injection every two minutes with shutters open). Use of the narrow-band rf BPMs and bending magnet x-ray BPMs together with two correctors per sector performs very well for the vertical plane.

Studies were performed using the bending magnet BPMs in a "local" configuration during user beam operation. For the bending magnet beamline in question, only one of the two available x-ray BPMs ws included in the algorithm, with an increased BPM weight in the matrix. The second unit, not included, was used to monitor the performance independently. In this case, stability was limited by the performance of the broadband rf BPMs straddling the source point, but was improved by an order of magnitude in comparison to operation without x-ray BPMs.

Using the "global" vertical corrector configuration (with as few broadband rf BPMs as possible) has allowed what is most likely the first true submicron beam stabilization with duration of more than five days. Shown in Figure 3 are data collected over a six-day period from two x-ray BPMs installed in an APS bending magnet beamline front end. Shown on the horizontal axis is the average of the two monitor readbacks, while the vertical axis shows the readback difference divided by longitudinal separation between the two units. Thus the vertical axis gives the x-ray beam's pointing angle in the vertical plane. Figure 3 is therefore a phase-space representation of beam stability in the frequency band extending from a fraction of a Hz down all the way to a five-day period. Admittedly, the horizontal axis is artificially small; after all, in this case both units were included in the orbit correction algorithm so it is not surprising that a small spread, less than half a micron rms, is observed. Even if one unit were to have systematic errors, the algorithm would simply track them in such a way as to minimize the average of the two units' readbacks. What is compelling, however, is that the rms vertical pointing angle over this five-day period (each day is represented by a different color dot) is 180 nanoradians. Thus, at a distance of 50 meters from the source, down along the x-ray beamline, the beam spot was stable to better than 10 microns rms over this five-day period. This is pretty darn good; I venture to say it is the best in the world.

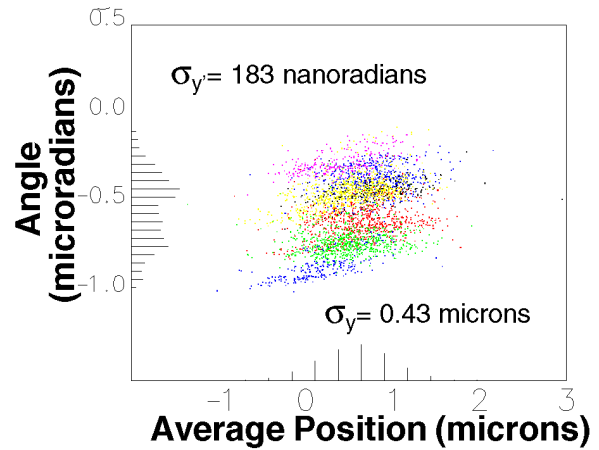

Figure 3. Phase-space representation of APS vertical beam position stability over a five-day period. Different colors correspond to separate days.

## 4 UPGRADE PLANS

While performance to date has been excellent in some areas, much remains to be done. Elimination of the 0.1-Hz system dead band described above will take the highest priority. Implementation of the 100-Hz data pool IOC for this purpose will allow further essential upgrades.

A major effort has been ongoing for the past two years to realign accelerator girders in order to reduce systematic errors in the ID x-ray BPMs associated with stray synchrotron radiation impinging on the photosensitive blades [9]. While this effort has done much to reduce background signals by a factor of between three and ten, it is clear that significant ID gap-dependent systematic errors remain, on the order of tens of microns. Research toward feedforward algorithms is promising, whereby a process monitors ID gaps and writes corrections to local steering correctors as the gap is changed in such a way as to reduce any residual steering error caused by the ID. In addition, this same process, or one like it, could make position offset corrections to the ID x-ray BPMs for that beamline, providing position readbacks that are relatively gap-independent. For this purpose, insertion device gap information will be placed on the reflective memory network, allowing real-time access by e.g., the new data pool IOC.

The present limitation on feedforward is the fact that in order to generate look up tables, one must carefully scan each of the over 22 IDs in series, with orbit correction running. Early tests show that, using existing algorithms, fully one hour of machine studies time is required for each ID, with only modest improvement, a factor of approximately three. By using the data pool, collection of feedforward data should take no longer than the time to configure the system and close the gap – one minute vs. one hour. Further, systematic drifts that creep in during the very slow measurement will be much reduced.

The real-time (1.5-kHz update rate) feedback system was completed by 1998 in its present incarnation [4]. It uses only the broadband rf BPMs to reduce beam noise in the band from 0.1 to 30 Hz. While new narrow-band and x-ray BPMs have become available since that time, they have been used exclusively for DC orbit control, running on a workstation in the main control room. This circumstance has mainly arisen from a limit in the area of applications programming. While narrow-band and x-ray BPM data are available to the 1.5-kHz system, the emphasis has been put on DC orbit control, which has been a difficult and high-priority challenge. Once the insertion device x-ray BPMs have been comfortably integrated into DC orbit control, including the new data pool IOC, attention will turn back to the 1.5-kHz real-time system. At present, orbit stability in the 0.1- to 30-Hz band is sitting around 1.3 microns rms vertically and 2.0 microns horizontally. As can be seen from Table 1, additional work in the real-time system is required if true submicron stability is to be realized. A long-range goal is to integrate all available BPM data (rf and x-ray) into the fast feedback algorithm.

## 5 CONCLUSIONS

Orbit correction technology at the APS is at a very advanced stage with the advent of new x-ray BPM data acquisition and processing. Long-term vertical pointing stability has been improved to better than 100 nanoradians rms for times scales from 10 seconds to 24 hours. Inclusion of all rf and x-ray BPM data on the reflective memory network will allow two orders of magnitude increase in the DC feedback update rate. This, in turn, will make possible feedforward algorithms to further eliminate beam motion transients induced by insertion device gap changes.

## 6 ACKNOWLEDGEMENTS

The real-time feedback system was conceptualized in the early 1990s by Youngjoo Chung. John Carwardine and Frank Lenkzsus corrected all the early conceptual errors and turned it into a working system [4]. Deming Shu performed the mechanical design of the x-ray BPMs in the tunnel.

This work was supported by the U.S. Department of Energy, Office of Basic Energy Sciences, under Contract No. W-31-109-ENG-38.